# $MoS_2$ nanoribbons as promising thermoelectric materials

D. D. Fan[1], H. J. Liu[1,*], L. Cheng[1], P. H. Jiang[1], J. Shi[1], X. F. Tang[2]

[1]*Key Laboratory of Artificial Micro- and Nano-Structures of Ministry of Education and School of Physics and Technology, Wuhan University, Wuhan 430072, China*

[2]*State Key Laboratory of Advanced Technology for Materials Synthesis and Processing, Wuhan University of Technology, Wuhan 430070, China*

The thermoelectric properties of $MoS_2$ armchair nanoribbons with different width are studied by using first-principles calculations and Boltzmann transport theory, where the relaxation time is predicted from deformation potential theory. Due to the dangling bonds at the armchair edge, there is obvious structure reconstruction of the nanoribbons which plays an important role in governing the electronic and transport properties. The investigated armchair nanoribbons are found to be semiconducting with indirect gaps, which exhibit interesting width-dependent oscillation behavior. The smaller gap of nanoribbon with width $N = 4$ leads to a much larger electrical conductivity at 300 K, which outweighs the relatively larger electronic thermal conductivity when compared with those of $N = 5, 6$. As a results, the room temperature *ZT* values can be optimized to 2.7 (*p*-type) and 2.0 (*n*-type), which significantly exceed the performance of most laboratory results reported in the literature.

Due to the increasing challenge of energy crisis and environmental pollution, searching for sustainable and clean energy has become more and more urgent. Thermoelectric materials which can directly convert heat into electricity and vice versa have attracted much attention from the science community. The efficiency of a thermoelectric material can be described by the dimensionless figure of merit $ZT = S^2\sigma T/(k_l + k_e)$, where $S$, $\sigma$, $T$, $k_l$, and $k_e$ are the Seebeck coefficient, the electrical conductivity, the absolute temperature, the lattice thermal conductivity and the electronic thermal conductivity, respectively. A good thermoelectric material has a large value of *ZT* which requires a large power factor ($S^2\sigma$) and/or low thermal conductivity ($k_l + k_e$). For conventional thermoelectric materials, these transport coefficients are usually coupled with each other, and it is generally difficult to significantly improve their thermoelectric performance. The situation becomes optimistic since the pioneering work of Hicks and Dresselhaus [1, 2], who found that the low-dimensional or nano-structured thermoelectric materials could exhibit much higher *ZT* values on account of the improved power factor caused by quantum confinement and energy filtering effects, as well as the reduced thermal conductivity because of the enhanced phonon boundary scattering. Although the experimental realization and characterization of such system remain a big challenge, it is still quite important to explore their thermoelectric performance theoretically, which may shed some light on searching new thermoelectric material with high efficiency.

---

[*] Author to whom correspondence should be addressed. Electronic mail: phlhj@whu.edu.cn

There is currently growing interests in the transition-metal dichalcogenide such as $MoS_2$ and its low-dimensional structures, which are believed to have wide application potentials in nano-electronics and optoelectronic devices [3, 4, 5]. However, the thermoelectric properties of $MoS_2$ and related structures are less investigated. Mansfield and Salam [6] measured the Seebeck coefficient of bulk $MoS_2$, which is about 600 $\mu V/K$ at room temperature and larger than those of most good thermoelectric materials. Buscema *et al*. [7] observed a very large and tunable Seebeck coefficient in monolayer $MoS_2$, which can be as high as $10^5$ $\mu V/K$ at low doping levels. Moreover, a low thermal conductivity of $MoS_2$ sheet is found both experimentally [8] and theoretically [9, 10]. Using first-principles calculations and Boltzmann transport theory, Guo *et al*. [11] predicted that the optimized *ZT* value of bulk $MoS_2$ is only 0.1 at about 700 K. Such poor thermoelectric performance can be attributed to a very low electrical conductivity of bulk $MoS_2$ [6, 12, 13]. In a further study, high pressure is applied to tune the inter-layer interactions of bulk $MoS_2$, and the *ZT* value can be increased to 0.65 over a wide pressure and temperature ranges [14]. Huang *et al*. [15] investigated the thermoelectric performance of $MoS_2$ monolayer by using two-dimensional ballistic transport model, and a highest *ZT* of about 0.58 is achieved for *p*-type doping at room temperature. Lee *et al*. [16]

performed first-principles calculations to study stacking of two different layers $MQ_1$ and $MQ_2$ (M = Mo, W, and $Q_1$, $Q_2$ = S, Se, Te), and they predicted that the mixed-layer compounds $MS_2/MTe_2$ can strongly enhance the thermoelectric properties as a consequence of reducing the band gap and the interlayer van der Waals interactions. Wickramaratne *et al*. [17] discussed the thermoelectric properties of few-layer $MoS_2$ within the framework of density functional theory, and found the maximum *ZT* value of 1.2 occurs for bilayer $MoS_2$. All these works suggest the possibility of using $MoS_2$ and its two-dimensional structures for thermoelectric applications, although the reported *ZT* values are not very high. Here we focus on the thermoelectric properties of one-dimensional $MoS_2$ nanoribbons, which is believed to exhibit a higher *ZT* than bulk and monolayer $MoS_2$ [1, 2]. Such kind of nanoribbons were previously made by electrochemical/chemical synthesis method [18, 19], with the width varying from tens of nanometers to hundreds of nanometers. Moreover, Wang *et al*. [20] synthesized $MoS_2$ nanoribbons encapsulated in CNTs, which have uniform widths down to 1~4 nm and layer numbers down to 1~3. Recently, $MoS_2$ nanoribbons with a uniform width of only 0.35 nm were widely formed between holes created in a $MoS_2$ sheet under electron irradiation [21]. In the present work, we consider the $MoS_2$ nanoribbons with different widths and our first-principles calculations indicate that the *ZT* value can be significantly enhanced to a value as high as 2.7 at room temperature, which significantly exceeds the performance of most laboratory results reported in the literature.

The electronic properties of $MoS_2$ nanoribbons are investigated by using first-principles project augmented wave (PAW) method as implemented in the Vienna *ab initio* simulation package (VASP) code [22, 23, 24]. The generalized gradient approximation (GGA) with Perdew-Bruke-Ernzerhof (PBE) functional [25] is used to calculate the exchange-correlation energy. We adopt a rectangular supercell where the nanoribbon and its periodic images are separated by a vacuum distance of at least 10 Å to avoid interactions. For the Brillouin zone integrations, we use a $3 \times 3 \times 9$ Monkhorst-Pack *k*-mesh scheme. The atomic positions are fully relaxed until the magnitude of the forces acting on all atoms becomes less than 0.01 eV/Å. Spin-orbit interactions are explicitly included in our calculations. Based on the calculated energy band structure, the Seebeck coefficient ($S$), the electrical conductivity ($\sigma$) are obtained by the semiclassical Boltzmann theory [26], where the relaxation time approximation is estimated from the deformation potential (DP) theory proposed by Bardeen and Shockley [27]. The electronic thermal conductivity $k_e$ is derived from the electrical conductivity $\sigma$ by using the Wiedemann-Franz law $k_e = L\sigma T$.

The MoS$_2$ nanoribbon structures can be viewed as tailoring a monolayer MoS$_2$ along the armchair or zigzag direction. Accordingly, the so-called armchair MoS$_2$ nanoribbons (AMNRs) and zigzag MoS$_2$ nanoribbons (ZMNRs) can be identified by the number of dimer lines or zigzag chains across the ribbon width and are labeled as *N*-AMNRs and *N*-ZMNRs, respectively. In this work, we focus on the armchair type and Fig. 1(a) shows the ball-and-stick model of *N*-AMNRs. We consider three different AMNRs with *N* = 4, 5, and 6, which correspond to a width of 4.72 Å, 6.62 Å, and 7.88 Å, respectively. Upon structural relaxations, the lattice constants along the extension direction are calculated to be 5.37 Å, 5.46 Å, and 5.46 Å, respectively. Remember that the translation distance in the same direction of monolayer MoS$_2$ is 5.54 Å, which indicates there exists obvious structure reconstruction when cutting the monolayer into nanoribbons. To clarify such reconstruction, we shown in Fig. 1(b) the relaxed structure of 6-AMNR as an example. We find that the distance between Mo1 and Mo2 atoms decreases from 3.20 Å to 3.01 Å upon geometry optimization. Moreover, the bond length of Mo1-S1 at the edge is decreased by 0.13 Å, while that near the center of the nanoribbon (Mo2-S2) is almost unchanged. The structure reconstruction is a result of dangling bonds at the edges, which will in turn have an important effect on the electronic and transport properties.

The band structures of three kinds of AMNRs are presented in Fig. 2. We see all of them are semiconducting, which is consistent with previous calculations for wider nanoribbons [28, 29, 30]. The calculated band gaps are indirect with values of 0.15, 0.49, and 0.44 eV for *N* = 4, 5, and 6, respectively. Note these values are much smaller than those found for the bulk (1.29 eV) [31] and monolayer MoS$_2$ (1.88 eV) [32], and can not be explained by the well-known quantum confinement effect. The reason is that the edge atoms of nanoribbons narrow the band gap by introducing new flat energy level at both conduction and valence band edges [28]. Additional calculations with more AMNRs included show that the variations of band gap as a function of ribbon width exhibit distinct oscillation behavior [37], and those ribbons with $N = 3p-1$ ($p$ is an integer) have larger band gaps than the neighboring ones. Such observation is similar to those found in the armchair graphene nanoribbons [33] and armchair silicon nanoribbons [34], and can be generalized as a robust characteristic of nanoribbons with armchair edges. As the band gap of 4-AMNR is much smaller than those of the other two nanoribbons, we believe they may exhibit quite different electronic transport properties. In addition, we find that the energy bands around the Fermi level are rather flat for all the investigated nanoribbons, which suggests a relatively larger effective mass and thus plays an important role in determining their transport properties.

Based on the calculated electronic band structures, the transport coefficients can be essentially derived by using the Boltzmann transport theory with the relaxation time approximation. In this method, Seebeck coefficient $S$ is independent of the relaxation time $\tau$, while the electrical conductivity $\sigma$ can only be calculated with respect to $\tau$. The accurate treatment of the relaxation time depends on the detailed scattering mechanism which is usually very complicated. As the wavelength of the thermally activated carriers at room temperature is much larger than the lattice constant and comparable to that of acoustic phonons, we believe the electron-acoustic phonon coupling dominates the scattering of carriers [35, 36], which can be effectively evaluated by the DP theory [27]. For one-dimensional systems, the relaxation time can be expressed as [37, 38, 39]:

$$\tau = \frac{\hbar^2 C_{1D}}{(2\pi k_B T)^{1/2} |m^*|^{1/2} E_1^2} \qquad (1)$$

Here $C_{1D} = \frac{1}{l_0} \frac{\partial^2 E}{\partial (\Delta l / l_0)^2}$ is the elastic constant, $m^* = \hbar^2 \left( \frac{\partial^2 E(k)}{\partial k^2} \right)^{-1}$ is the effective mass, and $E_1 = \frac{\partial E_{edge}}{\partial(\Delta l / l_0)}$ is the DP constant which represents the shift of band edges per unit strain. These three quantities can be readily obtained from first-principles calculations. The other parameters $e$, $\hbar$, $k_B$, and $T$ are the unit charge, the reduced Planck constant, the Boltzmann constant, and the absolute temperature, respectively. Our calculated results for the three kinds of AMNRs are summarized in Table I. We find that elastic constant and the absolute values of both effective mass and DP constant increase with increasing nanoribbon width, which is similar to that found in previous calculations for AMNRs with larger width [38]. Among the three investigated AMNRs, we find that the relaxation time of electrons and holes in 4-AMNR are both obviously larger than those in the other two nanoribbons, which originates from a smaller effective mass and DP constant. Moreover, the calculated relaxation time of 4-AMNR is also larger than that found in the monolayer $MoS_2$ [38]. All these suggest that $MoS_2$ nanoribbons with particular width could exhibit very favorable thermoelectric properties.

Another important factor that should be carefully treated is the Lorentz number $L$ when calculating the electronic thermal conductivity $k_e$ from the Wiedemann-Franz law. For most metallic systems, $L$ maintains a constant of $2.44 \times 10^{-8} W \cdot \Omega \cdot K^{-2}$ [40]. However, this is not the case for semiconductors especially at low doping level, where the Lorentz number is usually lower than $2.44 \times 10^{-8} W \cdot \Omega \cdot K^{-2}$. Depending on the reduced Fermi energy $\xi = E_f / k_B T$ and scattering parameter $r$, the Lorentz number can be expressed as [41]:

$$L = (\frac{k_B}{e})^2 \left( \frac{(r+7/2)F_{r+5/2}(\xi)}{(r+3/2)F_{r+1/2}(\xi)} - \left( \frac{(r+5/2)F_{r+3/2}(\xi)}{(r+3/2)F_{r+1/2}(\xi)} \right)^2 \right) \quad (2)$$

where
$$F_n(\xi) = \int_0^\infty \frac{\chi^n}{1+e^{\chi-\xi}} d\chi \quad (3)$$

As mentioned before, the acoustic phonon is the main scattering mechanism, and the scattering parameter is thus taken to be −0.5 [41].

With the relaxation time and Lorentz number available, we can now make a complete understanding of the electronic transport properties of MoS$_2$ nanoribbons. Fig. 3(a)-(c) plot the calculated room temperature Seebeck coefficient $S$, the electrical conductivity $\sigma$, and the electronic thermal conductivity $k_e$ as a function of chemical potential $\mu$. Within the rigid-band picture [42], the chemical potential corresponds to the doping level or carrier concentration of the system. For *p*-type doping $\mu$ is negative while it is positive for *n*-type doping. We see from Fig. 3(a) that the Seebeck coefficients of these AMNRs exhibit two obvious peaks around the Fermi level ($\mu = 0$). The absolute value is about 224 $\mu V/K$ for $N = 4$, and much larger for $N = 5, 6$ (~650 $\mu V/K$). Note the maximum Seebeck coefficient usually depends linearly on the band gap, which is larger for the case of $N = 5, 6$ than $N = 4$. Such an order is however reversed for the electrical conductivity shown in Fig. 3(b), where the 4-AMNR presents a much larger value compared to those found for 5-AMNR and 6-AMNR. This is consistent with the fact that 4-AMNR has a relatively larger relaxation time and smaller effective mass (see Table I). Note the electrical conductivity of 4-AMNR exhibits a small discontinuity at the Fermi level, which is caused by using different relaxation time for electrons and holes in our calculations. The electronic thermal conductivity $k_e$ shown in Fig. 3(c) almost coincides with the behavior of electrical conductivity, since $k_e$ is calculated from the Wiedemann-Franz law $k_e = L\sigma T$ where $L$ does not change too much according to our calculations.

To evaluate the *ZT* value, one also needs to know the lattice thermal conductivity $k_l$ of these MoS$_2$ nanoribbons. Here, we use the calculated result of Liu *et al.* [9] where the lattice thermal conductivity keeps a constant (1.02 W/mK) regardless of the nanoribbon width. Inserting all the transport coefficients into the expression of *ZT*, we can now predict the thermoelectric performance of MoS$_2$ nanoribbons. Fig. 3(d) shows the room temperature *ZT* values of three kinds of AMNRs. We see that by appropriately control the carrier concentration, the *ZT* value of 4-AMNR can be reached to 2.7 for *p*-type doping ($\mu = -0.027$ eV), and 2.0 for *n*-type doping ($\mu = 0.054$ eV). Such *ZT* values not only exceed those of bulk MoS$_2$ and its two-dimensional counterpart, but are also comparable to the best of those reported so far. In contrast, both the 5-AMNR and 6-AMNR exhibit relatively lower *ZT* values, which are respectively 1.0 and 0.8 for *p*-type doping, while 1.7 and 1.5 for *n*-type doping. We further find that the optimized *ZT* values of both *p*-type and *n*-type increase with decreasing ribbon width, which is consistent with those found in armchair graphene nanoribbons [33] and armchair silicon nanoribbons [34]. Once again, we observe a robust width dependent characteristic of armchair nanoribbons. Our calculated *ZT* values and the corresponding transport coefficients are summarized in Table II. We see that at the optimized carrier concentrations, all the three AMNRs actually have similar Seebeck coefficients with absolute value of about 200~300 $\mu V/K$. The significantly enhanced *ZT* values of 4-AMNR can be essentially attributed to its much higher electrical conductivity, which outweighs the relatively larger electronic thermal conductivity when compared with those of 5-AMNR and 6-AMNR.

In summary, our theoretical calculations demonstrate that MoS$_2$ nanoribbons with armchair edges could be optimized to exhibit very good thermoelectric performance. The predicted *ZT* values show obvious width dependence, and can be as high as 2.7 at room temperature, which is competitive to the best of those reported so far. Considering the fact that MoS$_2$ nanoribbons can be made by various synthesis techniques [18, 19, 20, 21], it is reasonable to expect that if the width and/or edge chirality can be experimentally controlled, MoS$_2$ nanoribbons could become very promising thermoelectric materials, which needs further theoretical and experimental investigations.

We thank financial support from the National Natural Science Foundation (Grant No. 51172167 and J1210061) and the "973 Program" of China (Grant No. 2013CB632502).

**Table I** The relaxation times at 300 K for *N*-AMNRs calculated by DP theory. The corresponding elastic constant, effective mass, and DP constant are also given.

| *N*-AMNRs | Carrier type | Elastic constant (eV/Å) | Effective mass ($m^*/m_e$) | DP constant (eV/strain) | Relaxation time(s) |
|---|---|---|---|---|---|
| *N* = 4 | electron | 24.2 | 1.55 | −1.26 | $5.55 \times 10^{-14}$ |
|  | hole | 24.2 | −1.38 | −0.96 | $1.02 \times 10^{-13}$ |
| *N* = 5 | electron | 35.2 | 1.79 | −2.0 | $2.97 \times 10^{-14}$ |
|  | hole | 35.2 | −1.42 | −3.0 | $1.48 \times 10^{-14}$ |
| *N* = 6 | electron | 48.7 | 2.07 | −2.89 | $1.78 \times 10^{-14}$ |
|  | hole | 48.7 | −2.41 | −3.95 | $9.12 \times 10^{-15}$ |

**Table II** Calculated room temperature *ZT* values at optimized chemical potentials (carrier concentrations) for *N*-AMNRs. The corresponding transport coefficients and energy band gaps are also listed.

| *N*-AMNRs | Gap (eV) | $\mu$ (eV) | $S$ ($\mu$V/K) | $\sigma$ (S/m) | $\kappa_e$ (W/mK) | $\kappa_l$ (W/mK) | *ZT* |
|---|---|---|---|---|---|---|---|
| *N* = 4 | 0.15 | −0.027 | 223 | $1.43 \times 10^6$ | 6.80 | 1.02 | 2.7 |
|  |  | 0.054 | −204 | $7.77 \times 10^6$ | 3.76 | 1.02 | 2.0 |
| *N* = 5 | 0.49 | −0.182 | 252 | $7.08 \times 10^4$ | 0.33 | 1.02 | 1.0 |
|  |  | 0.182 | −277 | $1.16 \times 10^5$ | 0.53 | 1.02 | 1.7 |
| *N* = 6 | 0.44 | −0.177 | 235 | $6.04 \times 10^4$ | 0.29 | 1.02 | 0.8 |
|  |  | 0.190 | −267 | $1.10 \times 10^5$ | 0.51 | 1.02 | 1.5 |

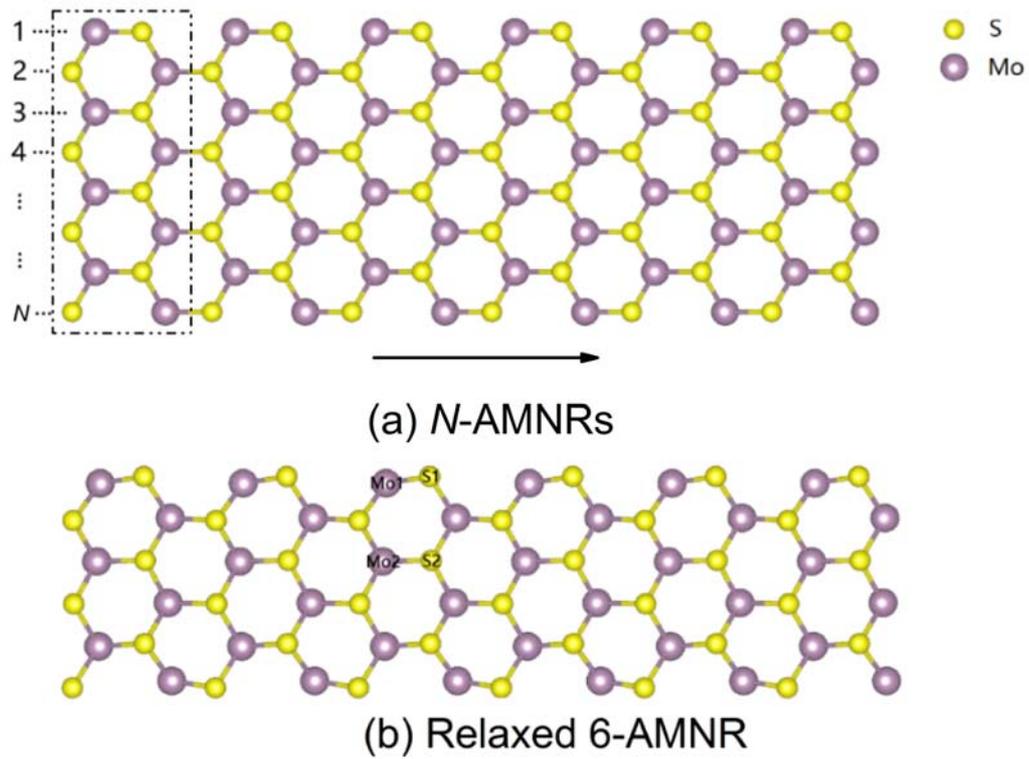

**Fig. 1** (Color online) Ball-and-stick model of (a) *N*-AMNRs, and (b) relaxed 6-AMNR. The arrow indicates the extension direction of nanoribbons, and the rectangle indicts the unit cell.

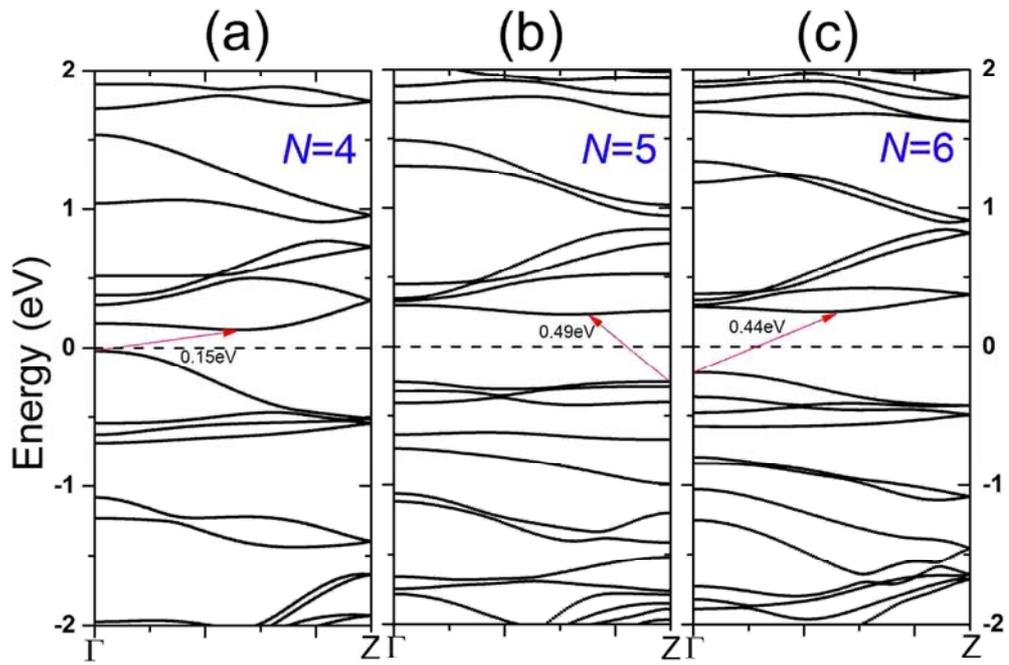

**Fig. 2** (Color online) Calculated energy band structures of *N*-AMNRs with *N* = 4, 5, 6. The Fermi level is at 0 eV and the band gaps are indicated by red arrows.

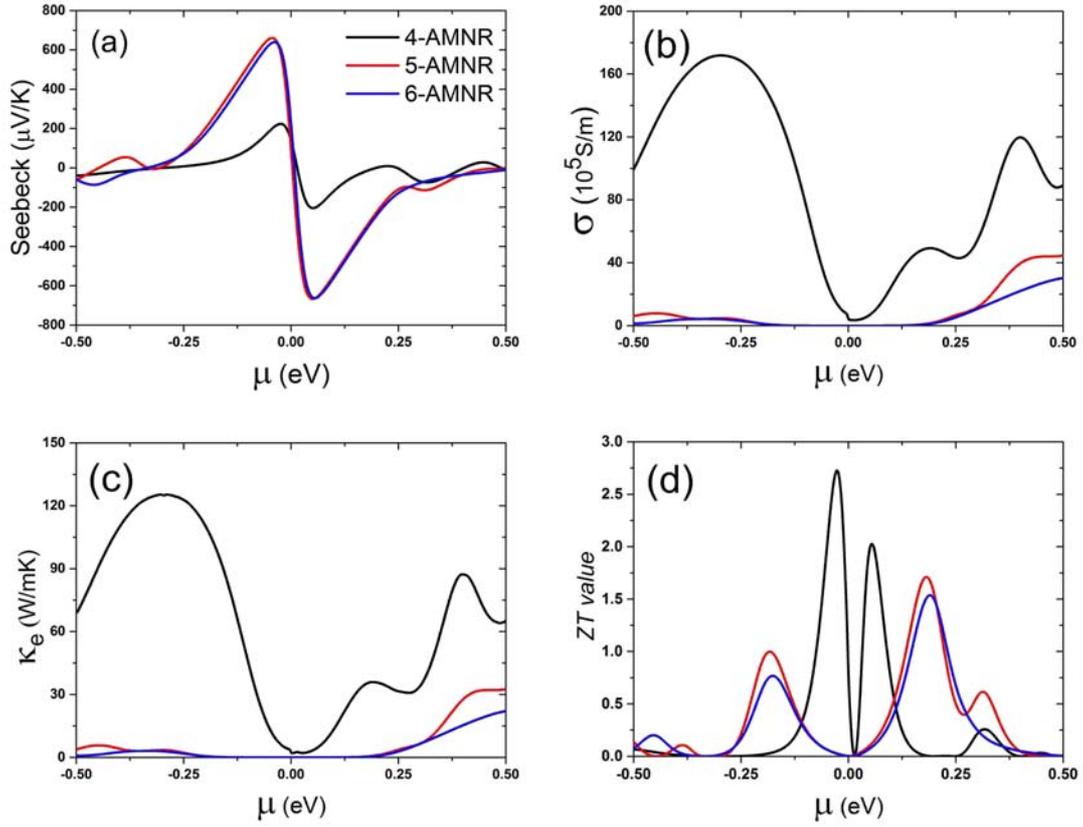

**Fig. 3** (Color online) Calculated electronic transport coefficients and *ZT* values as a function of chemical potential for *N*-AMNRs at 300 K: (a) Seebeck coefficient, (b) electrical conductivity, (c) electronic thermal conductivity, and (d) *ZT* value.